\documentclass{gGAF2e}
\def\ga{\mathrel{\mathchoice {\vcenter{\offinterlineskip\halign{\hfil
$\displaystyle##$\hfil\cr>\cr\sim\cr}}}
{\vcenter{\offinterlineskip\halign{\hfil$\textstyle##$\hfil\cr
>\cr\sim\cr}}}
{\vcenter{\offinterlineskip\halign{\hfil$\scriptstyle##$\hfil\cr
>\cr\sim\cr}}}
{\vcenter{\offinterlineskip\halign{\hfil$\scriptscriptstyle##$\hfil\cr
>\cr\sim\cr}}}}}

\def\la{\mathrel{\mathchoice {\vcenter{\offinterlineskip\halign{\hfil
$\displaystyle##$\hfil\cr<\cr\sim\cr}}}
{\vcenter{\offinterlineskip\halign{\hfil$\textstyle##$\hfil\cr  
<\cr\sim\cr}}}
{\vcenter{\offinterlineskip\halign{\hfil$\scriptstyle##$\hfil\cr
<\cr\sim\cr}}}
{\vcenter{\offinterlineskip\halign{\hfil$\scriptscriptstyle##$\hfil\cr
<\cr\sim\cr}}}}}

\begin{document}
\doi{10.1080/0309192YYxxxxxxxx}
\issn{1029-0419}
\issnp{0309-1929}
\jvol{00} \jnum{00} \jyear{2008} \jmonth{January}

\title{Spectral properties of oscillatory and non-oscillatory $\boldsymbol{\alpha^2}$-dynamos.}
\author{A. Giesecke\thanks{$^\ast$Corresponding
    author. Email: a.giesecke@hzdr.de \vspace{6pt}},
  F. Stefani, G. Gerbeth
  \\
  \vspace{6pt} {\em{Helmholtz-Zentrum Dresden-Rossendorf,
      Dresden, Germany}}}

\maketitle
\begin{abstract}
The eigenvalues and eigenfunctions of a linear $\alpha^2$-dynamo have
been computed for different spatial distributions of an isotropic
$\alpha$-effect.  Oscillatory solutions are obtained when $\alpha$
exhibits a sign change in the radial direction.  The time-dependent
solutions arise at so called exceptional points where two stationary
modes merge and continue as an oscillatory eigenfunction with
conjugate complex eigenvalues.  The close proximity of oscillatory and
non-oscillatory solutions may serve as the basic ingredient for
reversal models that describe abrupt polarity switches of a dipole
induced by noise.

Whereas the presence of an inner core with different magnetic
diffusivity has remarkable little impact on the character of the
dominating dynamo eigenmodes, the introduction of equatorial symmetry
breaking considerably changes the geometric character of the
solutions.  Around the dynamo threshold the leading modes correspond
to hemispherical dynamos even when the symmetry breaking is small.
This behavior can be explained by the approximate dipole-quadrupole
degeneration for the unperturbed problem.

More complicated scenarios may occur in case of more realistic
anisotropies of $\alpha$- and $\beta$-effect or through
non-linearities caused by the back-reaction of the magnetic field
(magnetic quenching).
\end{abstract}


\section{Introduction}
Oscillatory or reversing magnetic fields driven by a flow of an
electrically conducting fluid are a well known astrophysical
phenomenon.  Thus the solar magnetic field regularly oscillates on a
characteristic $22$ yrs time scale whereas the dominating dipole
component of the Earth's magnetic field irregularly changes its
orientation every few $100$ kyrs, conducting a so called reversal.
Sign changes of the magnetic field also have been observed in the
Cadarache von-K\'arm\'an-Sodium (VKS) dynamo, and depending on the
difference of the rotation rates of the two flow driving impellers
various regimes with oscillatory and/or chaotic behavior can be
obtained \citep{2007EL.....7759001B}.  Whereas the time scale of the
solar cycle can be reproduced using simple $\alpha\Omega$-dynamo
models, an explanation of the observed equatorial migration requires
further assumptions, e.g. meridional flow or negative radial shear at
the surface.  Regarding the geodynamo, three dimensional simulations
of the magnetohydrodynamic equations as well as mean field models have
been able to reproduce essential features of the Earth's magnetic
field \citep{2002PEPI..130..143H,2011PEPI..187..157C}, but others, for
example the very nature of the reversal mechanism or the large
variation of reversal rates \citep{1986PEPI...43...22M} still are
unsolved issues.  A prominent feature of this paleomagnetic reversal
frequency distribution is the occurrence of a few very long periods
($\ga 20$Myrs) during which no reversal occurred at all (so called
superchrons; \citeauthor{harlandbook}, \citeyear{harlandbook}).
Another surprising but less known property are deviations of the
distribution of inter-reversal time periods from an ideal Poisson
distribution \citep{2006PhRvL..96l8501C,2007PEPI..164..197S}
indicating long term correlations in the reversal trigger
mechanism(s).

In order to reliably disentangle internal reversal trigger mechanisms
(e.g. intrinsic changes of the fluid flow pattern in the liquid outer
core) from external sources (e.g. precession or changes in the heat
flux through the core mantle boundary) numerical simulations are
required that cover sufficient long time periods to allow for a large
number of reversal events.  In addition, numerical models should also
be capable to incorporate the mechanisms that trigger individual
reversals and reproduce general reversal characteristics such as
duration and field geometry during the actual reversal.  Mean field
models are ideally suited to execute long term simulations of the
(mean field) induction equation because they are computational cheap.
The price for this capability is a strong simplification by
parameterizing induction effects of the turbulent small scale flow
e.g. in terms of the $\alpha$-effect \citep{1980mfmd.book.....K}.
Hence, any internal changes of the small scale turbulence
(e.g. changes in statistical properties of the convection caused by
gradual growth of the solid inner core) are suspected to
encroach/expand into the comparably simple mean field coefficients.
Nevertheless, mean field modelling of reversing magnetic fields has
been remarkably successful
\citep{2002PEPI..130..143H,2005PhRvL..94r4506S,2009InvPr..25f5011F}.
The present work is based on the idea that a polarity change is part
of an oscillation of the dominant dipole mode
\citep{2007GApFD.101..227S}.  A consequential model for irregularly
occurring reversals requires a proximity of an oscillatory and a
non-oscillatory branch so that a single transition between both states
might be induced by noise.  This is substantiated by the fact that
polarity reversals in numerical dynamos are generally found in
intermediate parameter regions between stable (dipolar) dynamos with
small fluctuations and highly fluctuating and unstable dynamos
\citep{2006E&PSL.250..561O}.

The mean flow in the Earth's fluid core most probably is weak so that
usual geodynamo models are based on an $\alpha^2$ mechanism where an
$\alpha$-effect (re-)generates the poloidal field from the poloidal
field and vice versa.  A general requirement for the occurrence of
oscillating eigenmodes in $\alpha^2$-dynamos is $\nabla\alpha\neq 0$
\citep{1987AN....308..101R}.  However, for a long time it was assumed
that the leading eigenmode in simple $\alpha^2$-dynamos is
non-oscillatory and dominant oscillatory solutions are a curiosity
that requires a rather elaborate configuration
\citep{2003A&A...406...15R}.  Recently, it has been discovered that
oscillatory $\alpha^2$-dynamos are quite common when the radial
profile of a spherically symmetric, isotropic $\alpha$ exhibits a sign change
\citep{2003PhRvE..67b7302S} and a stringent mathematical treatment of
this model yields very general conditions for the occurrence of
oscillating solutions \citep{uwe_2010}.  Meanwhile oscillating
solutions have also been found in direct numerical simulations of
thermal convection in a spherical shell \citep{2011A&A...530A.140S} or
in a spherical wedge geometry with random helical forcing
\citep{2010ApJ...719L...1M}.  Such oscillating $\alpha^2$-dynamos
might provide an alternative approach for the solar dynamo, e.g. the
model of \citet{2010ApJ...719L...1M} exhibits equatorward migration
without the requirement of meridional circulation or negative radial
shear (which in the sun only occurs in a rather thin layer located
close to the surface).

Here, we examine the behavior of axisymmetric eigenmodes generated in
a two-dimensional kinematic $\alpha^2$-dynamo with isotropic
$\alpha$-effect.  The focus of our examinations is on the spectrum of
eigenvalues in terms of growth rates and oscillation frequencies for
two different radial profiles of the $\alpha$-effect.  For this
purpose we pick up the model of \cite{2005AN....326..693G} where a
sinusoidal radial $\alpha$ distribution is assumed.  We show that for
this $\alpha$-distribution the leading modes around the dynamo
threshold are oscillating independent of inner core conductivity or
latitudinal perturbations that would provide an equatorial symmetry
breaking.  In extension to \cite{2005AN....326..693G}, where only the
two dominating oscillating modes around the onset of dynamo action
were discussed, here we investigate the emergence/disappearance of
oscillating modes at so called exceptional points in more detail and
briefly identify the fundamental differences in the radial structure
of oscillatory and non-oscillatory eigenmodes.  The work aims at
building a bridge between original mean field models that have been
developed in the sixties of the last Century, and recent results from
observational and numerical approaches demonstrating the valuable
impact of mean field theory for the development of a geodynamo
reversal model.

\section{Equations and Method}
The basic assumption of the mean field approach is a separation of the
velocity field $\bm{u}$ and the magnetic field $\bm{B}$ which are
split into mean parts $\left<\bm{u}\right>$, $\left<\bm{B}\right>$
and fluctuating parts $\bm{u}'$, $\bm{B}'$ according to
\begin{equation}
\bm{B} = \left<\bm{B}\right> + \bm{B'}\mbox{ and } \bm{u} =
\left<\bm{u}\right> + \bm{u'},
\label{eq::2scale}\end{equation}
where $\left< \cdot \right>$ represents an appropriate space- and time
average so that the Reynolds averaging rules apply.  Then the temporal
development of the mean magnetic field $\left<\bm{B}\right>$ is
described by the mean field induction equation
\begin{equation} 
\partial_t\left<\bm{B}\right>={\bm{\nabla}}\times\left(\left<\bm{u}\right>\times\left<\bm{B}\right> 
+\bm{\mathcal{E}}-{\eta}{\bm{\nabla}}\times\left<\bm{B}\right>\right)
\label{eq::ind}
\end{equation}
with the turbulent electromotive force (EMF)
$\bm{\mathcal{E}}=\left<\bm{u'}\times{\bm{B'}}\right>$ that
describes the average induction action of (unresolved) small scale
flow perturbations.  $\bm{\mathcal{E}}$ results from the interaction
of the fluctuating velocity field $\bm{u}'$ with the mean magnetic
field $\left<\bm{B}\right>$ and is formally represented by a linear
functional of $\left<\bm{u}\right>$, $\left<\bm{B}\right>$ and the
statistical properties of $\bm{u}'$ \citep{1980mfmd.book.....K}.
Assuming that the mean quantities vary only slightly around a certain
space-time point, only contribution from a certain neighborhood must
be taken into account and the EMF can be written as a Taylor
expansion:
\begin{equation}
\mathcal{E}_i=\left<\bm{u'}\times{\bm{B'}}\right>_i=\alpha_{ij}\left<B_j\right>
+\beta_{ijk}\frac{\partial\left<B_k\right>}{\partial x_j}+...
\end{equation}
where the coefficients $\alpha_{ij}$ and $\beta_{ijk}$ depend on the
properties of the turbulent fluctuations $\bm{u}'$.  In general, the
turbulence is anisotropic so that $\boldsymbol{\alpha}$- and
$\boldsymbol{\beta}$-effect are described by tensors of 2nd and 3rd
rank, respectively.  Here we restrict ourselves to the simplest case
of isotropic turbulence so that $\alpha$ and $\beta$ are given by
\begin{eqnarray}
{\alpha}_{ij} & = & \alpha_0\delta_{ij}, \\ {\beta}_{ijk} & = &
-\beta\epsilon_{ijk}
\end{eqnarray}
so that $\alpha$ and $\beta$ are prescribed by scalar quantities.
Note that for a non-vanishing $\alpha$-effect the turbulence
additionally must be non-mirrorsymmetric.  With the definition of the
turbulent diffusivity
\begin{equation}
\eta_{\rm{T}}=\eta+\beta
\end{equation}
and vanishing mean flow $\left<\bm{u}\right>=0$ the mean field
induction equation simplifies to an equation describing the temporal
development of an $\alpha^2$-dynamo:
\begin{equation}
\partial_t\bm{B}={\bm{\nabla}}\times\left(\alpha\bm{B}
-\eta_{\mathrm{T}}{\bm{\nabla}}\times\bm{B}\right)
\label{eq::alphasq}
\end{equation}
(the brackets around the mean field $\bm{B}$ from now on are dropped
for simplicity).  Equation~(\ref{eq::alphasq}) is linear in $\bm{B}$
and the ansatz $\bm{B}=\bm{B}(\bm{r})e^{\gamma t}$ leads to a
linear eigenvalue problem
\begin{equation}
\mathcal{M}\bm{B}=\gamma\bm{B}\label{eq::eigenvalue}
\end{equation}
with the matrix $\mathcal{M}$ containing the dynamo operator from the
right hand side of~(\ref{eq::alphasq}), and the eigenvalue
$\gamma=\sigma+i\nu$ (growth rate $\sigma$ and frequency $\nu$).  For
uniform distributions of $\alpha$ and $\eta_{\rm{T}}$ (semi-) analytic
solutions are known (e.g. \citeauthor{1980mfmd.book.....K},
\citeyear{1980mfmd.book.....K}).  For more complex flows and spatial
distributions of $\alpha$ and $\eta$ a numerical solution of the
eigenvalue problem is required (e.g. \citeauthor{1972A&A....18..453R}
\citeyear{1972A&A....18..453R}, \citeauthor{1989RSPSA.425..407D}
\citeyear{1989RSPSA.425..407D}).  Here we solve~(\ref{eq::eigenvalue})
in a sphere surrounded by a non-conducting vacuum applying the method
presented in \cite{2010A&A...519A..80S}.  The approach is based on the
biorthogonality of the electric current
$\bm{j}=\mu_0^{-1}{\bm{\nabla}}\times\bm{B}$ and the vector potential $\bm{A}$
and explicitly utilizes an expansion of the magnetic field $\bm{B}$
into (analytical known) free decay modes.  The method allows a very
fast computation of eigenvalues and eigenfunctions for nearly
arbitrary spatial distributions of $\alpha$ and/or $\eta_{\rm{T}}$.
Further we restrict our examination to the axisymmetric field which is
dominating for isotropic $\alpha$ and $\beta$. Higher azimuthal modes
will be important in case of anisotropic $\alpha$ coefficient
\citep{2004GApFD..98..225T}, although in case of anisotropic
$\eta_{\rm{T}}$ axisymmetric modes become again dominant
\citep{2007AN....328.1130E}.

\section{Results}
\begin{figure}[t!]
\includegraphics[width=\textwidth]{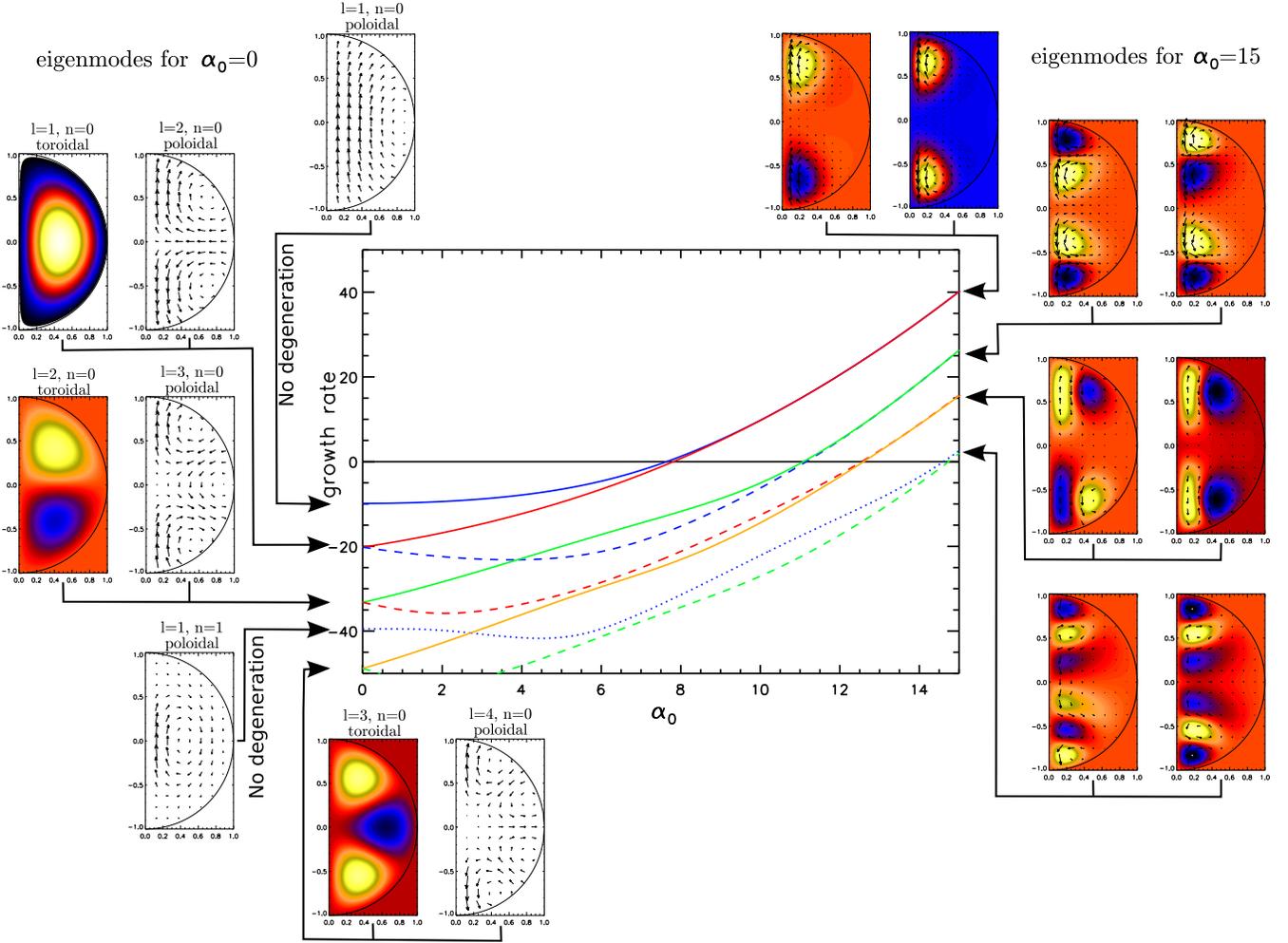}
\caption{(colour online) Central panel: growth rates for an $\alpha^2$ dynamo with
  $\alpha=\alpha_0\cos\vartheta$. Note the transition between
  different types of degeneration
  when going from $\alpha_0=0$ (free decay) to $\alpha_0\gg
  \alpha_{\rm{crit}}$. The small panels
  surrounding the central figure present the geometric structure of the
  eigenmodes for $\alpha_0=0$ (left hand side) and for $\alpha_0=15$ (right
  hand side). The colored contours denote the toroidal part and the
  arrows denote the poloidal part. Note the separation of toroidal and
  poloidal components in the free decay case ($\alpha_0=0$, left). 
}\label{fig::alpha_01}
\end{figure}

We start with a full sphere (embedded in vacuum) where $\alpha$ is
created from turbulent motions in a convective layer subject to
rotation. Then the EMF contains a term parallel to the mean magnetic
field   $\bm{\mathcal{E}} \propto (\bm{g}{\bm{\cdot}}\bm{\Omega})\bm{B}$ 
where $\bm{g}$ points in the radial direction and $\bm{\Omega}$ is
parallel to the rotation axis ($z$-axis) so that
$(\bm{g}{\bm{\cdot}}\bm{\Omega})\propto\cos\vartheta$.   
Thus, the $\alpha$-effect
is maximum at the poles and vanishes at the equator:
\begin{equation}
\alpha =\alpha_0\cos\vartheta.
\end{equation}
This $\alpha$ model has been examined e.g. in
\cite{1972RSPTA.272..663R}. Here we revive these results and
additionally present eigenvalues and eigenfunctions of higher order
modes. 
The resulting spectrum of the leading dynamo eigenmodes in terms of
the growth rates versus the amplitude $\alpha_0$ is
shown in figure~\ref{fig::alpha_01}.

All eigenvalues are real, i.e. all eigenfunctions are non-oscillatory modes. 
Due to the strict antisymmetry of $\alpha$ with respect to the equator 
the parity of the eigenmodes remains preserved for increasing $\alpha$. 
That is, the $\alpha$-term in the induction equation
couples toroidal and poloidal modes in a way that the dynamo
eigenmode remains antisymmetric (in the following we call this
"dipolar-like") or symmetric (in the following we call this
"quadrupolar-like") with respect to the equator. 
At $\alpha_0=0$ (free decay) an exact degeneration between "consecutive"
modes is obtained (as indicated on the left side of figure~\ref{fig::alpha_01}). 
The degenerated eigenfunctions are always a pair of a purely toroidal
and a purely poloidal mode which differ in multipolar degree by $\Delta
l=1$. 
The degeneration vanishes for $\alpha_0 > 0$ and around the dynamo
threshold a slight predominance of the dipole
mode occurs. 
The critical $\alpha$ for the onset of dynamo action of the dipole
mode ($\alpha_0^{\rm{crit}}=7.645$) is close to the value for the
quadrupolar mode ($\alpha_0^{\rm{crit}}=7.813$). Both values agree
with the results obtained by \cite{1972RSPTA.272..663R}
($\alpha_0^{\rm{crit}}=7.637$ for the dipole mode and
$\alpha_0^{\rm{crit}}=7.803$ for the quadrupole mode).
For $\alpha \gg \alpha_{\rm{crit}}$ all eigenfunctions again approach a twofold
degenerated state that consists of a pair of 
degenerated modes with similar
geometric structure, but opposite equatorial symmetry (see right side in
figure~\ref{fig::alpha_01}). 
The approximate degeneration between dipolar-like and quadrupolar-like modes
is the result of a symmetry in the equations describing the
$\alpha^2$-dynamo which is exact in case of perfect conducting
boundary conditions.
In that case dipolar eigenfunctions and quadrupolar eigenfunctions are
adjoints of each other and thus have the same (conjugate)
eigenspectrum \citep{1977AN....298...19P}.  
The degeneration is approximately retained in case of insulating
boundary conditions so that the eigenvalues remain close to each other
\citep{1998GApFD..89...45D}.

\subsubsection*{Radial dependence of $\alpha$}
\begin{figure}[h!]
\begin{center}
\includegraphics[width=8cm]{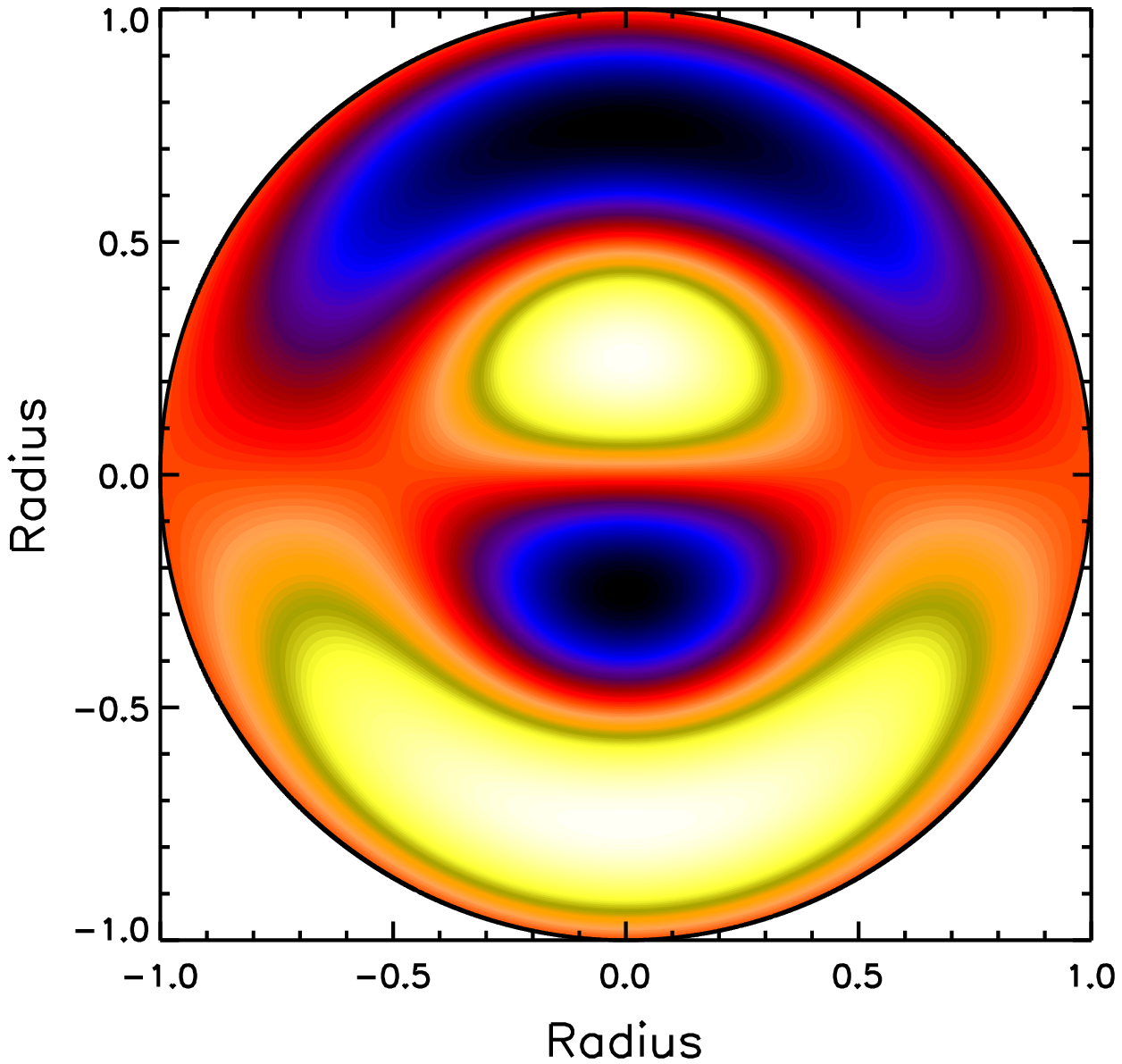}
\includegraphics[width=8cm]{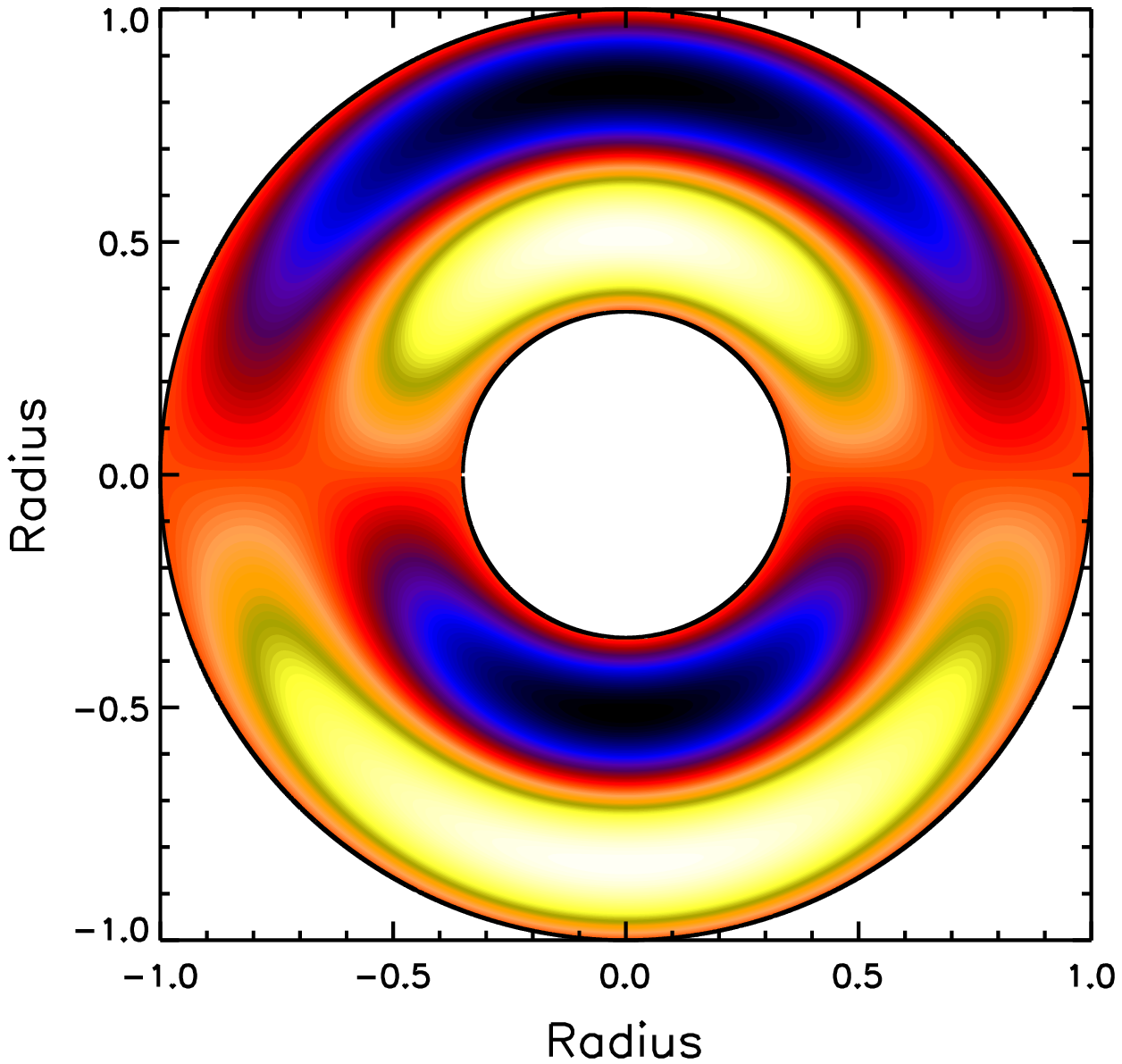}
\end{center}
\caption{(colour online) Radial and latitudinal distribution of the $\alpha$-effect as
given by~(\ref{eq::geo_alpha}). Left: $R_{\rm{in}}=0$; Right:
  $R_{\rm{in}}=0.35$ corresponding to the actual size of the Earth's
  inner core.}\label{fig::alpha_pattern}
\end{figure}

Next, we introduce a radial dependence for $\alpha$ taken from the
model presented in \cite{2005AN....326..693G}. This model utilizes an
idealized parameterization in the radial direction that is based on   
local simulations of rotating magnetoconvection assuming conditions roughly
suitable for the  geodynamo (weak stratification, fast rotation,
strong magnetic field; \citeauthor{2005PEPI..152...90G},
\citeyear{2005PEPI..152...90G}). 
Qualitatively these simulations show that
on the northern hemisphere $\alpha$ is
negative in the lower part of the liquid outer
core and positive in the upper part, whereby the zero occurs roughly
in the middle of the convective instable layer. 
The idealized radial $\alpha$-profile that incorporates these properties
is given by
\begin{equation}
\alpha(\bm{r})=\alpha_0\cos\vartheta\sin
\left(2\pi\frac{(r-R_{\rm{in}})}{(R_{\rm{out}}-R_{\rm{in}})}\right), 
\label{eq::geo_alpha}   
\end{equation}
where the outer radius of the spherical domain is fixed to
${R_{\rm{out}}}=1$ and $R_{\rm{in}}$ represents the radius of an inner
core (see figure~\ref{fig::alpha_pattern}).
A change in the radial profile was
also considered in the first planetary mean field models
\citep{1969AN....291..271S}. 
However, in that work $\alpha$ was related
to the radial derivative of the turbulence intensity $u_{\rm{rms}}^2$ and
the authors exclusively looked for non-oscillating solutions.

\begin{figure}[h!t]
\includegraphics[width=9cm]{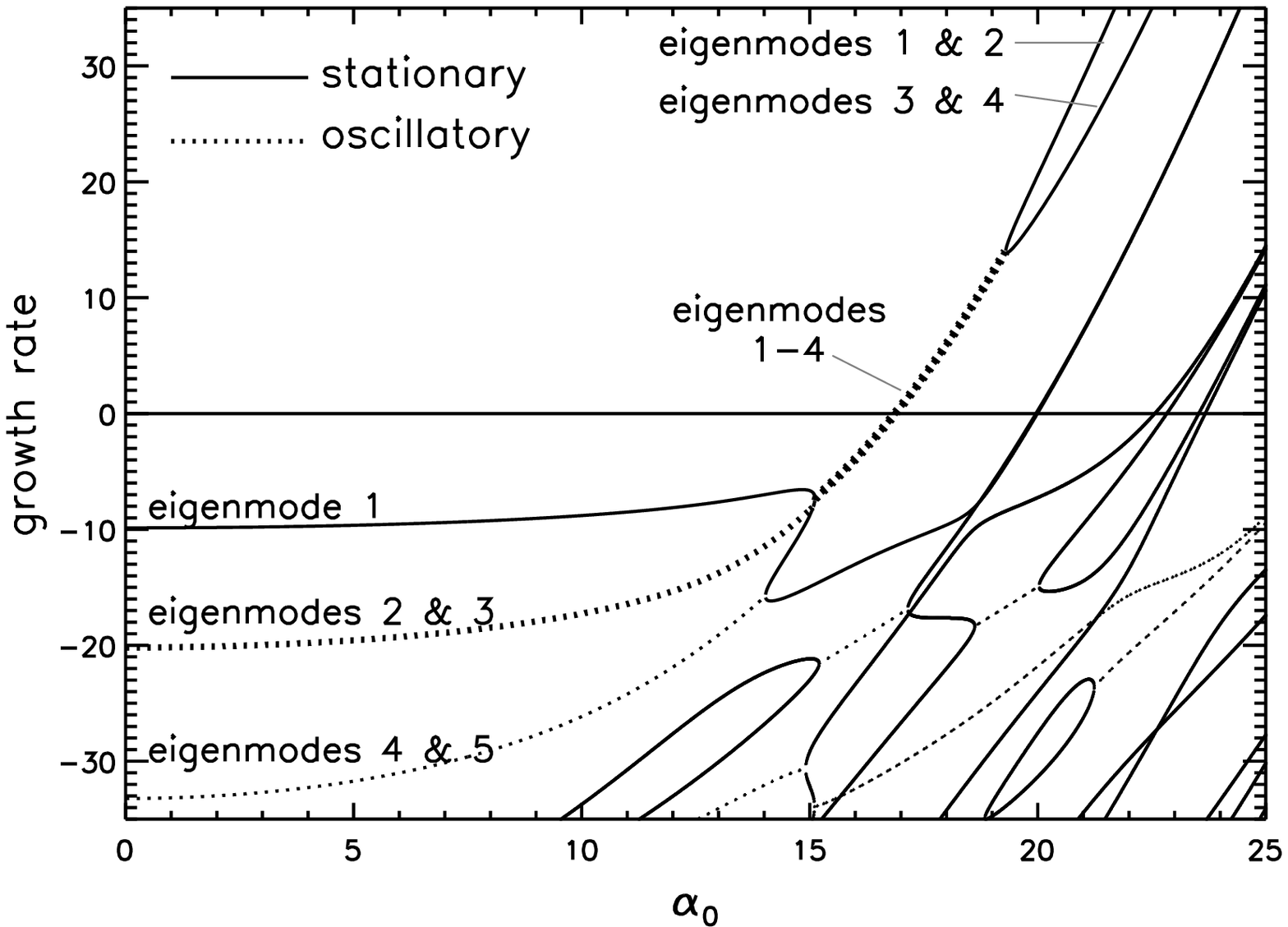}
\nolinebreak[4!]
\includegraphics[width=9cm]{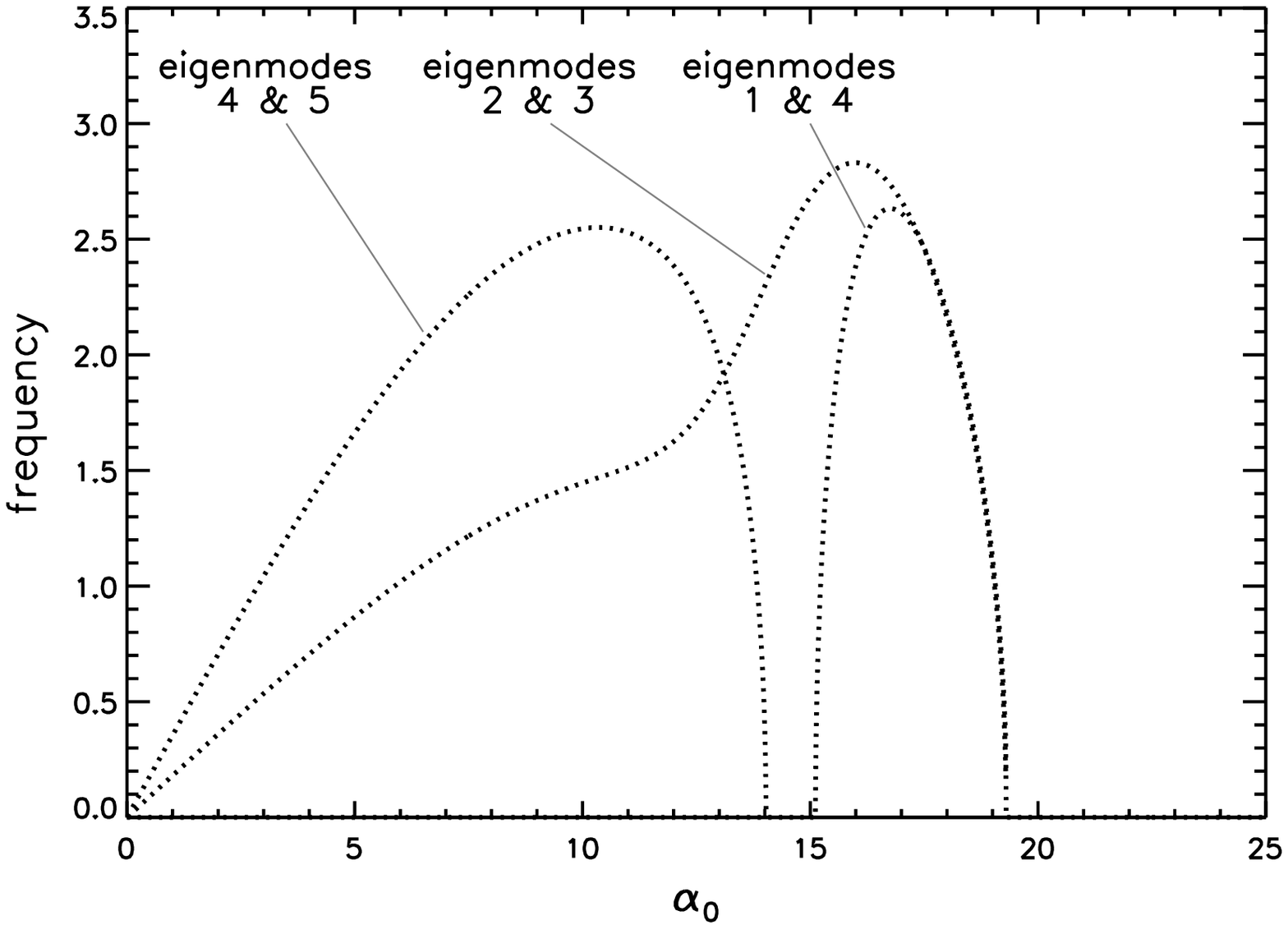}
\caption{Left panel: growth rates versus $\alpha_0$. The leading
  eigenmodes are numbered in the order of their appearance at
  $\alpha_0=0$. 
  Right
  panel: oscillation frequency versus $\alpha_0$ (only
  the positive frequencies for the three leading time dependent modes 
  are shown).} 
\label{fig::spectrum_sinr}
\end{figure}

Here, we start assuming no inner core, i.e. $R_{\rm{in}}=0$.
The spectrum for the leading axisymmetric modes is
shown in figure~\ref{fig::spectrum_sinr}. 
In contrast to the simple $\alpha$-profile examined in the previous
section, oscillatory solutions (indicated by dotted curves) are
obtained for this new radial $\alpha$ profile. 
In particular, around the dynamo threshold the dominating eigenmodes
are oscillatory dipolar-like (blue curve) and quadrupolar-like (red
curve) solutions.
The oscillating solutions appear/disappear at so called exceptional
points (EP) where two stationary modes coalesce and continue as two
oscillatory modes with conjugate complex eigenvalues, i.e. with the
same growth rate and positive and negative frequency, respectively
(the right panel of figure~\ref{fig::spectrum_sinr} shows only the
positive frequency branch).    
Regarding the appearance of oscillatory modes at
$\alpha=0$ (dotted curves on the left panel of figure~\ref{fig::spectrum_sinr}) it
is necessary to stress that exactly at $\alpha =0$ (i.e. for free
decay) these modes are stationary and degenerated, i.e. $\alpha=0$ is
an exceptional point for these modes.
The spectrum is far more complex than in models without any radial
structure of the $\alpha$-effect and which is apparent by the
confusing structure of the eigenvalues with a variety of level
crossings and/or exceptional points where  
oscillatory solutions constitute or vanish. 
However, for $\alpha\gg\alpha_{\rm{crit}}$ predominantly stationary modes are
observed.

\subsubsection*{Decomposition in free decay modes}
\begin{figure}[h!]
\includegraphics[width=9cm]{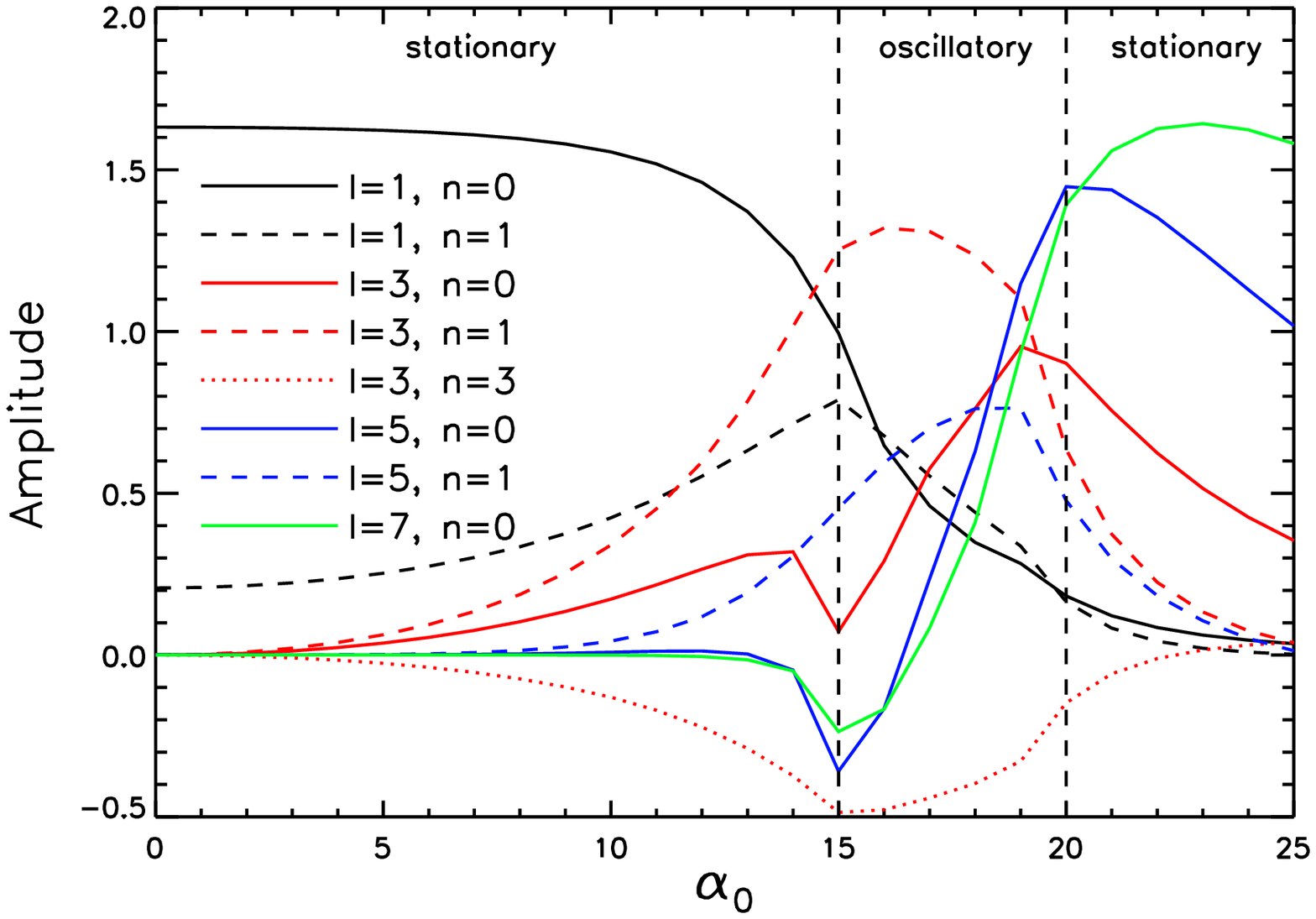}
\nolinebreak[4!]
\includegraphics[width=9cm]{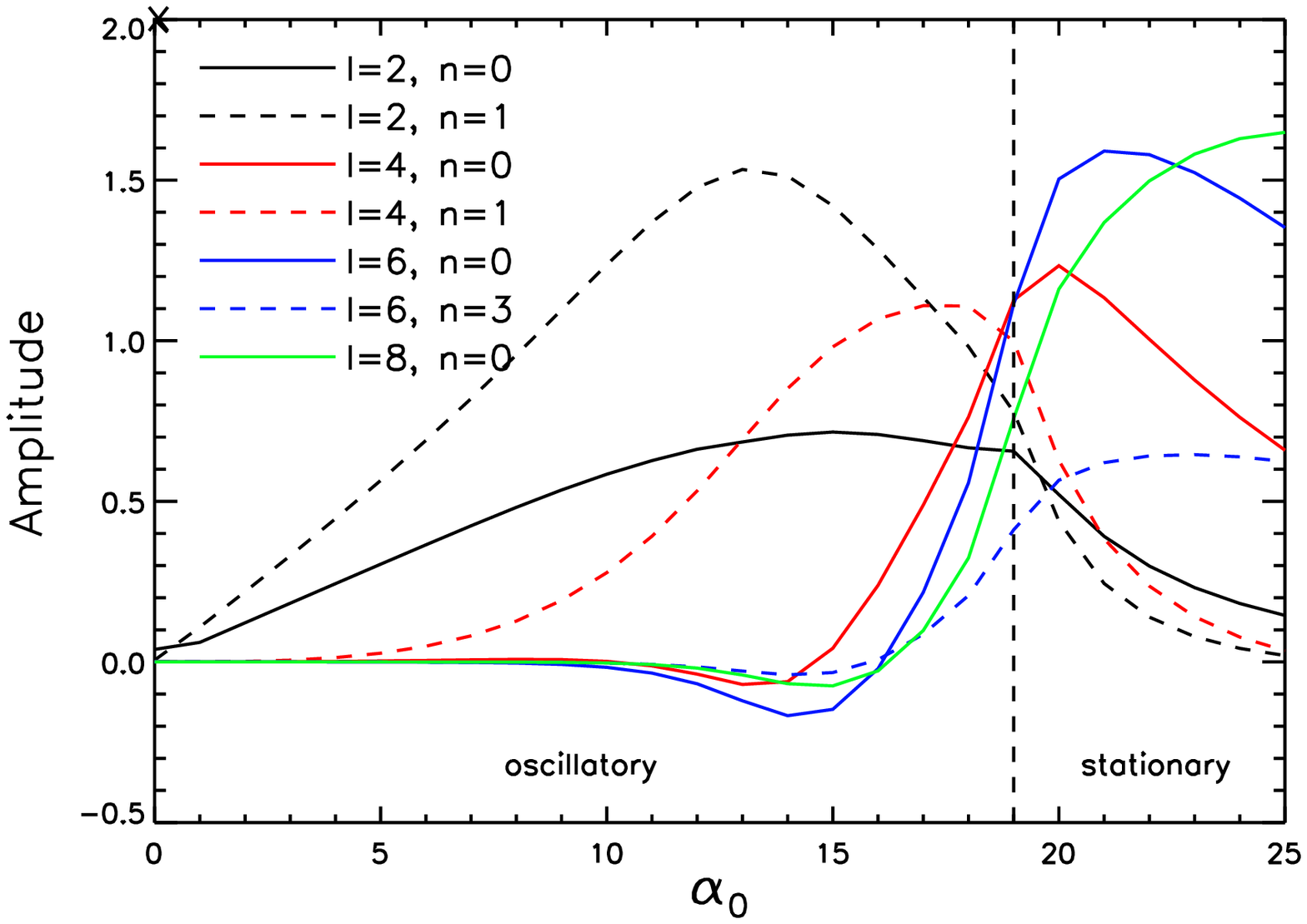}
\caption{Contribution of characteristic free decay modes to the first
  dipolar-like dynamo eigenmode (left panel, eigenmode 1 in
  figure~\ref{fig::spectrum_sinr}) and the first quadrupolar-like 
  dynamo eigenmode (right panel, eigenmode 2 \& 3 in figure~\ref{fig::spectrum_sinr}).}
\label{fig::decomposition_sinr}
\end{figure}

The change of the temporal character of the dynamo eigenmodes from
non-oscillatory to oscillatory behavior is attended by a change in the
radial structure. 
This can be seen by means of a
decomposition of a dynamo eigenmode in terms of the leading free
decay modes (which also represent the basis utilized in the numerical
scheme).
The dominating contributions in dependence on the magnitude of the
$\alpha$-effect are shown in figure~\ref{fig::decomposition_sinr}. 
For both symmetry classes the oscillatory regime is characterized by
domination of modes with radial wavenumber $n=1$ whereas stationary
solutions are characterized by a domination of
$(n=0)$ modes\footnote{$n$ characterizes the degree of a spherical
  Bessel function of a particular free decay mode and determines the
  radial behavior.}.
Another feature shown in the decomposition of the dynamo eigenmode is
the increment of the multipolar degree of the dominant contribution
with increasing $\alpha$.

\subsubsection*{Inner core}
The influence of an inner core with finite electrical conductivity had
been examined by \cite{1993Natur.365..541H,1995PEPI...87..171H} and
\cite{2002PEPI..132..281W} with surprisingly opposite
conclusions. 
Whereas the latter only found little influence of finite inner core
conductivity on temporal behavior of the dipole field,
\cite{1993Natur.365..541H, 1995PEPI...87..171H} concluded from
simulations that the finite conductivity of an inner core has a
stabilizing impact and reversals could only occur if the field in the
fluid outer core exhibits a large 
and long lasting fluctuation that allows the field to reverse
throughout the inner core as well.

\begin{figure}[h!]
\includegraphics[width=6cm]{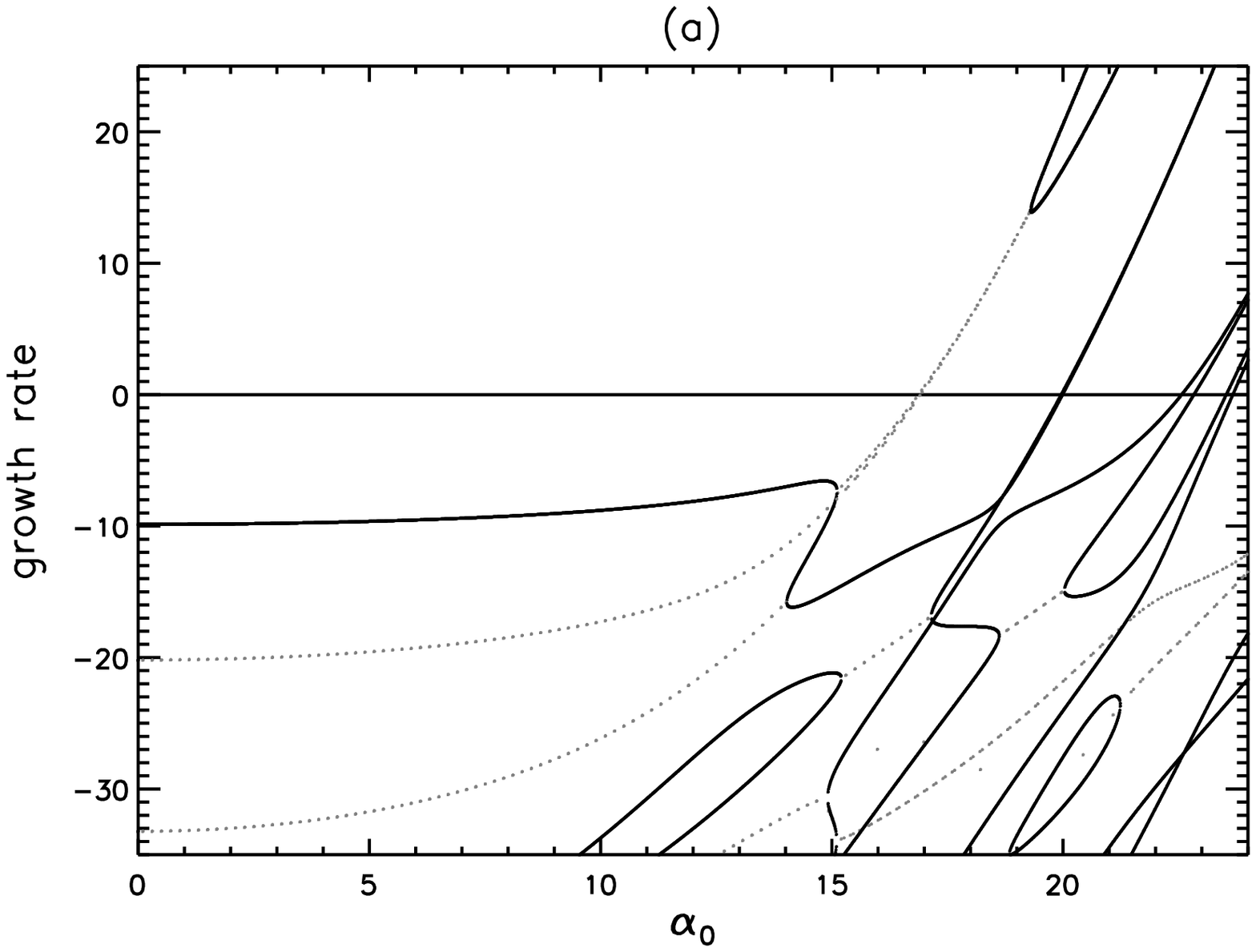}
\hspace*{-0.2cm}
\includegraphics[width=6cm]{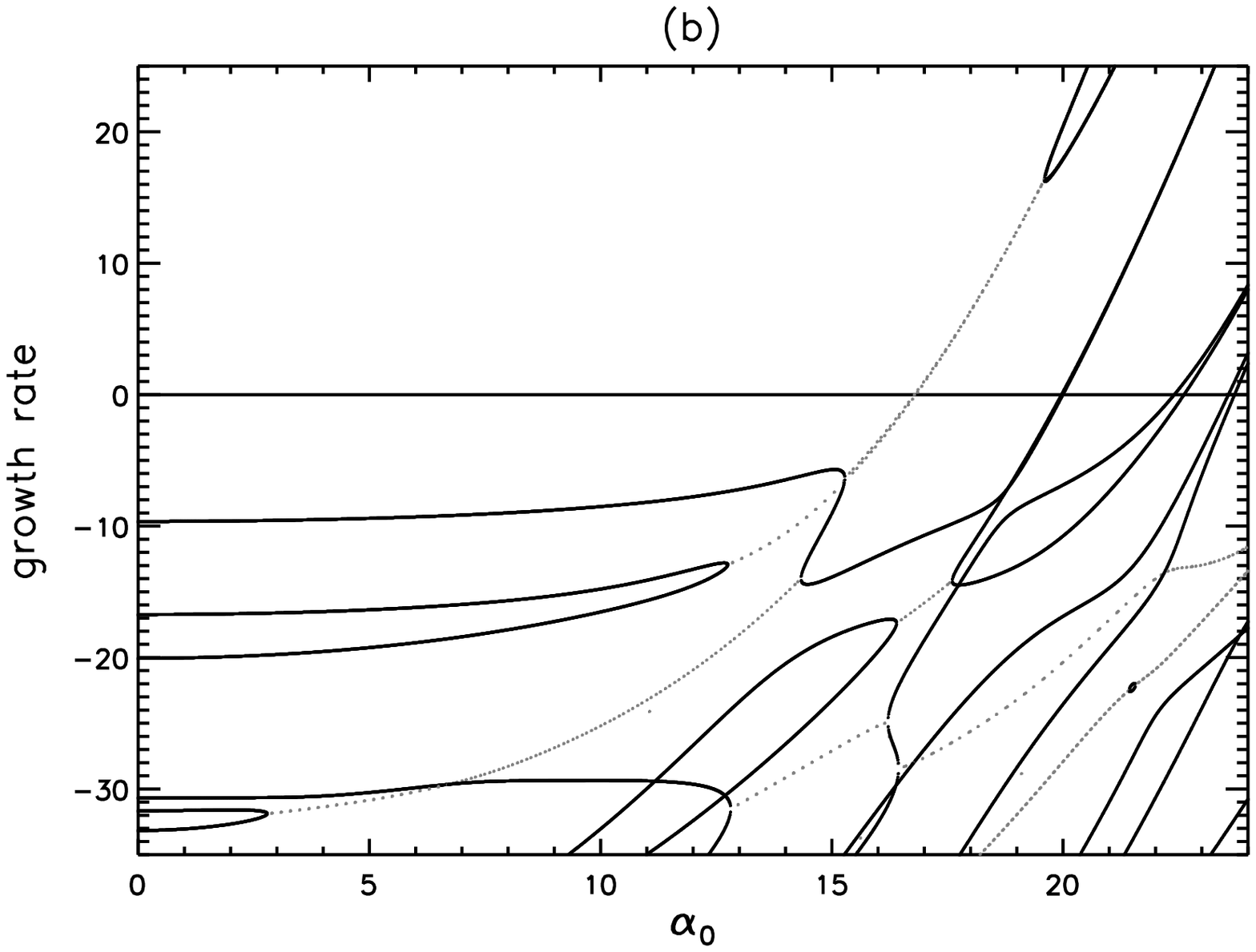}
\hspace*{-0.2cm}
\includegraphics[width=6cm]{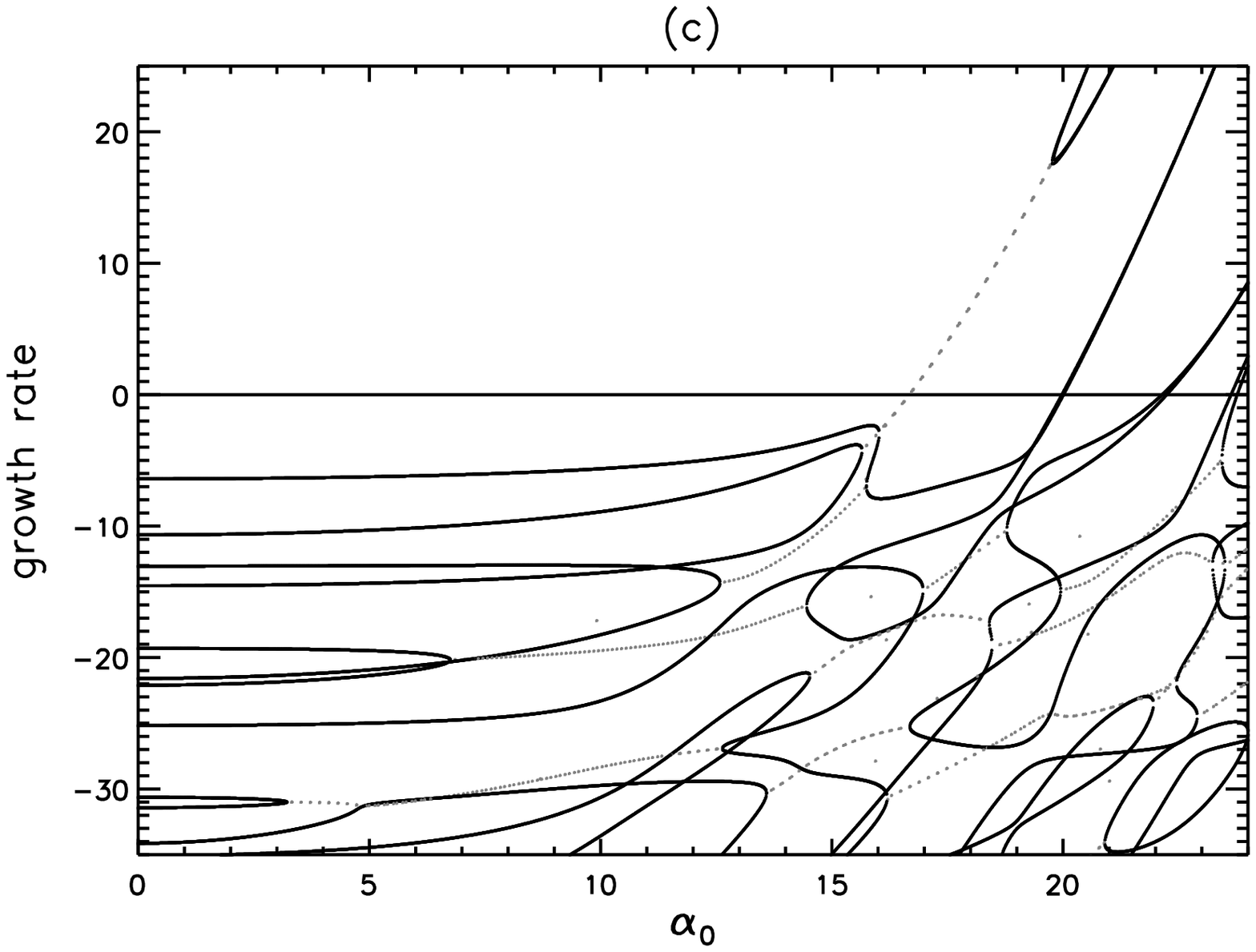}
\caption{Growth rates versus $\alpha_0$ for a modified $\alpha$-profile
  given by~(\ref{eq::geo_alpha}). An inner core is considered with an
  (earth-like) radius ${R_{\rm{in}}}=0.35$ with a magnetic diffusivity
  $\eta_{\rm{core}}=1, 0.5, 0.1$ (from left to right). Solid black curves
  represent non-oscillatory solutions and grey/dotted curves show
  oscillatory solutions.}\label{fig::gr_vs_alphasin_ic}
\end{figure}

Figure~\ref{fig::gr_vs_alphasin_ic}(a) shows that the spectrum of the
eigenmodes is only slightly changed when an inner core is considered
with a radius ${R_{\rm{in}}=0.35}$ and a uniform diffusivity within
inner and outer core ($\eta_{\rm{T}}=1$).  The $\alpha$-effect is
prescribed by the distribution~(\ref{eq::geo_alpha}) with $\alpha=0$
for $r<R_{\rm{in}}$.  The results for the same size of the inner core,
but with a reduced diffusivity for $r<R_{\rm{in}}$, is shown in
figures~\ref{fig::gr_vs_alphasin_ic}(b) and (c).  The introduction of
an inner core without changing the diffusivity distribution has
remarkable little influence on the spectrum
(figure~\ref{fig::gr_vs_alphasin_ic}a) whereas a reduced inner core
diffusivity for sufficiently small $\alpha$ results in a split-up of
the oscillatory modes into two stationary branches
(figure~\ref{fig::gr_vs_alphasin_ic}b and c).
However, around the onset of dynamo action the typical pattern of the
leading dynamo eigenmodes is hardly changed with increasing inner core
conductivity.  For example, the reduction from $\eta_{\rm{core}}=1$ to
$\eta_{\rm{core}}=0.1$ leads only to a small decrease of the critical
$\alpha$-magnitude from $\alpha_{\rm{crit}}\approx 16.9$ to
$\alpha_{\rm{crit}}\approx 16.5$.  Furthermore, the temporal behavior
of the leading modes is not changed within this regime, i.e. the
leading modes remain oscillating around the onset of dynamo action
independent of the inner core conductivity.  In particular for
$\alpha\gg \alpha_{\rm{crit}}$ no changes can be observed in the
behavior of the leading modes.  The main impact of an enhanced inner
core conductivity is manifested in the cancellation of the
degeneration of the eigenfunction for $\alpha\la \alpha_{\rm{crit}}$.
Consequently, the spectrum becomes more complex than in case of no
inner core or with uniform conductivity.

\subsubsection*{Equatorial symmetry breaking}
A coupling between dipole and quadrupole is the basis of the low
dimensional reversal model of
\cite{2008JPCM...20W4203P,2009PhRvL.102n4503P} where the authors 
conclude that the reversal rate is constrained by breaking of the
equatorial symmetry. 
With this motivation in the background, we investigate the influence
of equatorial symmetry breaking of the $\alpha$-effect. In the model
an additional term is added proportional to $\cos(2\vartheta)$ leading to
\begin{equation}
\alpha=\alpha_0\sin\left(\frac{2\pi\left(r-R_{\rm{in}}\right)}{R_{\rm{out}}-R_{\rm{in}}}\right)  
(\cos\vartheta+b\cos(2\vartheta))
\end{equation}
where $b\ll 1$ and again $R_{\rm{in}}=0.35$. 
The main impact of the parameter $b$ is a coupling of dipolar-like and
quadrupolar-like modes so that a distinction in terms of parity with
respect to the equator is no longer possible.

\begin{figure}[h!t]
\includegraphics[width=\textwidth]{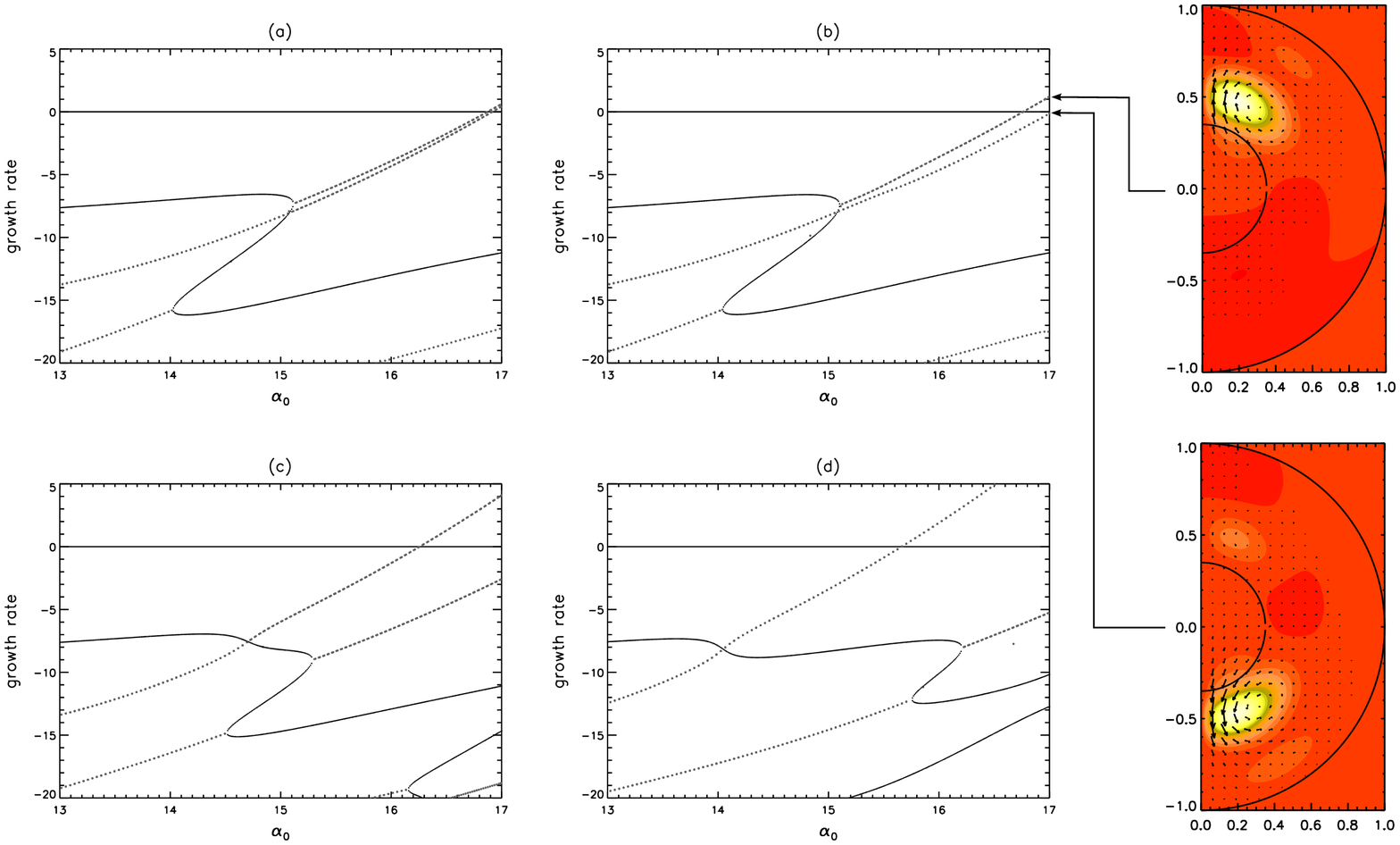}
\caption{Extraction of growth rates vs. $\alpha_0$ for dynamo models with
  equatorial symmetry breaking demonstrating the impact of symmetry
  breaking on the degeneration of the eigenfunctions. Solid curves
  denote stationary solutions and dotted curves denote oscillatory
  solutions. $b=0.00, 0.01, 0.05, 0.10$.  The contour plots on the
  right hand side show the pattern of the (hemispherical) eigenmodes
  for case (b) at $\alpha_0=17$ (colour online).
\label{fig::gr_eqsymbreak}}
\end{figure}

The resulting growth rates for various values of the parameter $b$ are
shown in figure~\ref{fig::gr_eqsymbreak} whereby $\alpha$ is
restricted to values around the dynamo threshold.
Without symmetry breaking two oscillatory eigenfunctions that
correspond to an oscillatory dipolar-like mode and an oscillatory
quadrupolar-like mode cross the dynamo threshold around $\alpha_0\approx
16.9$.  The coupling of dipolar-like and quadrupolar-like modes
results in a number of significant changes in the eigenvalues and
geometric structure.  Here we restrict the discussion to the behavior
around the dynamo threshold of the leading eigenmode.  First of all,
we observe a reduction of the critical $\alpha$ with increasing $b$
which is attended by a breakup of the degeneration of the leading
eigenmodes.  However, the dominating eigenmodes retain their
oscillatory property but due to the coupling of both symmetry classes
a distinction into dipolar-like and quadrupolar-like eigenmodes is no
longer possible.  Compared to the undisturbed case, a considerable
change in the field structure occurs for sufficiently strong $\alpha$.
Around the dynamo threshold (and above) the coupling of both classes
results in the occurrence of hemispherical dynamos that oscillate in
time.  Since dipolar-like and quadrupolar-like modes contribute
roughly the same amount to the new coupled mode (recall that the
growth rates of both types are roughly equal in the case without any
perturbation) the coupled dynamo mode is concentrated within one
hemisphere.  Since dipolar and quadrupolar modes have a very similar
structure, their contribution cancel out in one hemisphere so that
magnetic energy is concentrated in the remaining hemisphere.

\section{Conclusions}
We have re-confirmed in a simple mean-field dynamo model that
oscillatory $\alpha^2$-dynamos are possible when the radial profile of
$\alpha$ exhibits a more complex structure.  Our model is based on a
particular radial profile of the $\alpha$-effect that behaves $\propto
\sin(r)$.  The choice of this profile is not arbitrary but is
motivated by simulations of rotating magnetoconvection
\citep{2005PEPI..152...90G} and quasi-linear computations performed by
\citet{1974RSPTA.275..611S}.  On a first glance the occurrence of
oscillatory solutions seems quite robust.  Around the onset of dynamo
action, growing oscillatory solutions dominate independently of core
conductivity (and/or core size) or equatorial symmetry breaking.  More
important for a disappearance of oscillatory solutions might be
deviations from the ideal $\sin$-profile such as shifts of the zero
crossing \citep{2005AN....326..693G} which has not been examined here.

The eigenfunctions of the $\alpha^2$-model exhibit a couple of
remarkable properties that characterize the geometric structure of the
eigenfields.  In the oscillatory regime the eigenfunctions are
dominated by contributions with higher radial wavenumber whereas the
steady solutions essentially are determined by eigenfunctions with
sparse radial structure.  Hence, it is suggestive to look for an
indication of a similar contribution of higher radial modes during an
reversal in three-dimensional MHD simulations of the geodynamo or in
field reconstructions from paleomagnetic observations as e.g. executed
by \cite{2007E&PSL.253..172L}.

For $\alpha$-distributions with perfect equatorial (anti-)symmetry
$\alpha\propto\cos\vartheta$ the dipolar and quadrupolar modes remain
separated since no interaction between these modes is possible.  The
eigenmodes with different equatorial symmetry exhibit an approximate
degeneration which would be exact in case of perfectly conducting
boundary conditions.  The corresponding proximity of growth rates for
dipole and quadrupole has a dramatic impact when a small perturbation
is considered that breaks the equatorial symmetry.  In the vicinity of
the dynamo threshold the resulting coupling between dipolar- and
quadrupolar-like modes leads to hemispherical dynamo action.  Indeed,
the spatial reconstruction of the last reversal (the Matuyama-Brunhes
transition $\sim 780$ kyrs ago) shows a growing contribution of the
quadrupolar component during the actual reversal
\citep{2007E&PSL.253..172L} but there are no hints for hemispherical
dynamo action taken place in the Earth's core.  Nevertheless,
hemispherical dynamo action might have been the reason for the
non-uniform crust magnetization as a result of the ancient martian
dynamo \citep{2011PEPI..185...61L}.

Our results differ from the achievement of \cite{2009PhRvE..80c5302G}
who examined a kinematic $\alpha^2$ model using a strongly localized
$\alpha$-effect concentrated in two thin shells.  In their model it is
the coupling between dipole and quadrupole induced by equatorial
symmetry breaking which leads to an oscillating eigenmode, whereas in
our model the coupling between different radial modes (induced by the
radial variation of $\alpha$) is responsible for the oscillatory
behavior.  The simplicity of both models, however, prevents a robust
conclusion which specific behavior indeed is realized in the
geodynamo.

Regarding the Earth's magnetic field, a reason for the suppression of
the quadrupole might arise from anisotropies of the turbulent flow
essentially caused by the fast rotation of the Earth.  These
anisotropies resulting from the fast rotation of the Earth which
suppresses variations/fluctuations along the rotation axis should be
reflected in more realistic models of the $\alpha$-tensor as well as
in the diffusivity tensor.  Using more realistic anisotropic structure
of the $\alpha$-tensor (but isotropic $\eta$) results in domination of
non-axisymmetric (i.e. $m=1$) modes \citep{2005AN....326..693G}.  This
domination can be circumvented by assuming anisotropic description for
$\eta$ as well \citep{2004GApFD..98..225T}.  Note that
non-axisymmetric modes indeed seem to be important in the
reconstruction of the field pattern during the last reversal (the
Matuyama-Brunhes transition) where the most important field
contribution is determined by a $m=2$ contribution
\citep{2007E&PSL.253..172L}.

A substantial limitation of the presented results is the linear
character of the kinematic models, in particular regarding the
proximity of the eigenvalues of leading dipolar and quadrupolar
eigenmodes.  Thus a robust prediction on the behavior in a saturated
state is difficult and more elaborated mean field models will require
the consideration of backreaction of the magnetic field by virtue of
$\alpha$-quenching.  Qualitatively, we expect a similar behavior as it
has been described by \cite{2005PhRvL..94r4506S} in a one-dimensional
non-linear model of a reversing $\alpha^2$-dynamo.  Assuming that the
magnetic quenching of the $\alpha$-effect determines the instantaneous
growth rate the authors showed that the corresponding eigenmode is
inevitably driven towards the oscillatory branch. This shift is
attended by self accelerating field decay and the emerging local
maximum of the growth rate in connection with noise provide the
essential preconditions to switch from a stationary branch to an
oscillatory branch and vice versa.

\section*{Acknowledgments}
Financial support from Deutsche Forschungsgemeinschaft (DFG) in frame of the
Collaborative Research Center (SFB) 609 is gratefully acknowledged.
%
\bibliographystyle{gGAF} 
\bibliography{giesecke_osc_alpha2} 
\end{document}